\documentclass[useAMS,usenatbib]{mn2e}
\usepackage{graphicx}
\usepackage{amssymb}
\usepackage{natbib}
\usepackage{longtable}
\usepackage[T1]{fontenc}
\usepackage{color}

\title[Point source detection in WMAP data]{Extragalactic point source
  detection in WMAP 7-year data at 61 and 94 GHz}

\author[Lanz et al.]{
\parbox[t]{\textwidth}
{L.~F. Lanz$^{1,2}$\thanks{E-mail: lanz@ifca.unican.es},
D. Herranz$^{1,3}$, M. L\'opez-Caniego$^{1}$,  
J. Gonz\'alez-Nuevo$^{4}$ , G. de Zotti$^{5,4}$,
M. Massardi$^{5,6}$ and J.~L. Sanz$^{1}$} \\
\vspace*{8pt} \\
$^{1}$ Instituto de F{\'\i}sica de Cantabria, CSIC-UC, Av. de Los Castros s/n, Santander, 39005, Spain \\
$^{2}$ Departamento de F{\'\i}sica Moderna, Universidad de Cantabria, Av. de Los Castros s/n, Santander, 39005, Spain \\
$^{3}$ Astrophysics Group, Cavendish Laboratory, J.J. Thomson Avenue, Cambridge CB3 0HE, UK \\
$^{4}$ SISSA, via Bonomea 265, I-34136 Trieste, Italy \\
$^{5}$ INAF - Osservatorio Astronomico di Padova, Vicolo dell'Osservatorio 5, I-35122 Padova, Italy \\
$^{6}$ INAF - Istituto di Radioastronomia, Via P. Gobetti 101, I-40129 Bologna, Italy }

\begin{document}

\date{Received --, Accepted --}

\pagerange{\pageref{firstpage}--\pageref{lastpage}} \pubyear{2012}

\maketitle

\label{firstpage}

\begin{abstract}

  The detection of point sources in Cosmic Microwave Background maps
  is usually based on a single-frequency approach, whereby maps at
  each frequency are filtered separately and the spectral information
  on the sources is derived combining the results at the different
  frequencies. On the contrary, in the case of multi-frequency
  detection methods, source detection and spectral information are
  tightly interconnected in order to increase the source detection
  efficiency.

  In this work we apply the \emph{matched multifilter} method to the
  detection of point sources in the WMAP 7yr data at 61 and 94
  GHz. This linear filtering technique takes into account the spatial
  and the cross-power spectrum information at the same time using the
  spectral behaviour of the sources without making any a priori
  assumption about it. We follow a two-step approach. First, we do a
  blind detection of the sources over the whole sky. Second, we do a
  refined local analysis at their positions to improve the
  signal-to-noise ratio of the detections. At 94 GHz we detect 129
  $5\sigma$ objects at $|b|>5^\circ$ (excluding the Large Magellanic
  Cloud region); 119 of them are reliable extragalactic sources and
  104 of these 119 lie outside the WMAP Point Source Catalog
  mask. Nine of the total 129 detections are known Galactic sources or
  lie in regions of intense Galactic emission and one additional
  (weak) high-Galactic latitude source has no counterpart in
  low-frequency radio catalogues. Our results constitute a substantial
  improvement over the NEWPS-3year catalogue.

\end{abstract}


\section{Introduction} \label{introduccion}

The study of the Cosmic Microwave Background (CMB) provides a very
useful tool to understand the Universe and its evolution. A proper
analysis of this primordial radiation allows us to discriminate
between different evolutionary models of the Universe. Different
experiments have taken data with different observing conditions
(i.e. resolutions, frequencies, fields of view, ...) in order to
improve our comprehension of this radiation and, therefore, of the
Universe. The excellent sensitivities of modern instruments, close to
fundamental limits, imply that our ability to accurately measure the
CMB temperature anisotropies is limited by the contamination of CMB
maps by astrophysical emissions (foregrounds).  For this reason, many
techniques have been developed in order to separate the different
components that one can find in the maps. While on moderate to large
angular scales the main contaminants are diffuse Galactic emissions,
extragalactic point sources dominate on small angular scales both in
temperature~\citep{tof98,zotti99,hob99,zotti05} and in
polarisation~\citep{tucci04,tucci05,powps,argueso11b}.

Astrophysical foregrounds can be both a disturbance for CMB studies
and interesting {\it per se}. The WMAP surveys have made possible to
investigate for the first time the statistical properties of bright
radio sources above $\sim 10\,$GHz over the whole sky
\citep{wmap0,hinshaw07,NEWPS07,chen08,gnuevo08,massardi09,wright09short,wmap7yr}. The
first \emph{Planck} results \citep{ERCSC} have already offered the
possibility of extending the study to higher frequencies and fainter
flux densities \citep{ERCSCprop}.

There is a rich literature on methods to extract compact sources from
CMB maps \citep[see][for a recent review]{herranz_vielva}. Successful
techniques exploit a variety of tools:
wavelets~\citep{vielva01,vielva03,MHW2006,wsphere,NEWPS07}, matched
filters~\citep{vikhlinin95,tegmark98,barreiro03,can06} and other
linear filtering
tools~\citep{sanz01,naselsky02,herr02c,can04a,can05a,can05b}. Usually
these methods filter maps at each frequency separately. In addition,
Bayesian techniques that include prior information about the
distribution of the sources have been proposed in the
literature~\citep{hob03,psnakesI,argueso11a}.


Although single-frequency filtering techniques have been remarkably
successful, a multi-frequency approach allows us to take advantage of
additional information such as the cross-power spectrum of the noise.
This makes possible to improve the significance of sources seen at
different frequencies but with relatively low signal-to-noise ratio in
each frequency channel without any a priori assumption on their
spectral properties. However, multi-frequency detection of point
sources in CMB maps is still a poorly explored
field. \cite{herranz08a} introduced the matrix filters technique
(MTXF) as the first fully multi-frequency, non-parametric, linear
filtering technique that is able to find point sources and to do
unbiased estimations of their flux densities. The MTXF technique
exploits the distinctive spatial behaviour of these sources without
assuming any specific spectral behaviour. \cite{mtxf09} applied the
MTXF to realistic simulations of the \emph{Planck} radio channels,
showing that it is possible to practically double the number of
detections for some of the channels with respect to the
single-frequency matched filter approach for a fixed reliability
level.

The MTXF approach exploits the multi-frequency information only on the
diffuse components (like CMB, Galactic emissions and noise). A more
complete approach should take into account also the correlations among
the different frequency maps due to point sources themselves. A step
in this direction was done by \cite{herr02a} who presented the matched
multi-filter (MMF) method, a generalisation of the standard matched
filter for multi-frequency data where the frequency dependence of the
signal is known. This is the case for the thermal Sunyaev-Zel'dovich
(SZ) effect \citep{SZ70,SZ72}, for which the method was originally
developed and that has been used to build the Planck Early SZ
catalog. Unfortunately this is not the case for extragalactic radio
sources. However, \cite{lanz10} took advantage of an analogy between
the SZ and the point source case. In the SZ case, the spectral
behaviour of the sources is known but the size of the clusters is
not. To deal with this, the size of the source is described, in the
design of an MMF, by means of a free parameter, whose value is
determined maximising the signal-to-noise ratio for each detected
source \citep{herr02a}. In the case of point sources, the angular
profiles are known (they are given by the point spread functions,
PSFs, of the instrument) and we only need to optimise a set of
parameters describing the spectral shape.

In \cite{lanz10} the MMF was applied to realistic simulations of the
\emph{Planck} mission. The promising results obtained in that work
prompted us to apply the method to real data, namely to the 7-yr WMAP
data. More precisely, we have considered the V and W maps (61 and 94
GHz respectively), because at these frequencies the knowledge of the
statistical properties of radio galaxies was poor. The structure of
the article is as follows. In \S\,\ref{metodo} we briefly describe the
method, and in \S\,\ref{datos} the data used. In \S\,~\ref{resultados}
we present and discuss our results, comparing our flux density
estimates with measurements with other instruments. Our main
conclusions are summarised in \S\,\ref{conclusiones}.

\section{Method} \label{metodo}

The MMF is the optimal linear detection method when the spatial
profile and the frequency dependence of the sources are known. By
`optimal' we mean, as it is common in statistics, that the estimation
of the flux density of the sources is unbiased and has minimum
variance (maximum efficiency). In Fourier space the MMF takes the
form:
\begin{equation} \label{eq:mmf}
\mathbf{\Psi}(q)=\alpha \ {\bf P}^{-1} \mathbf{F}, \ \ \ \
\alpha^{-1}=\int{d{\bf q} \ \mathbf{F}^T{\bf
P}^{-1} \mathbf{F}},
\end{equation}
\noindent
where $q$ is the Fourier mode, $\mathbf{\Psi}(q)$ is the column vector
of the filters $\mathbf{\Psi}(q)=[\psi_{\nu}(q)]$, $\mathbf{F}$ is the
column vector $\mathbf{F}=[f_{\nu}\tau_{\nu}]$, $f_{\nu}$ is the
frequency dependence, $\tau_{\nu}(q)$ is the source profile at each
frequency $\nu$ and ${\bf P}^{-1}$ is the inverse matrix of the
cross-power spectrum {\bf P}. The MMF takes as arguments $N$ images
(in this work $N=2$) and returns a single filtered image where the
sources are optimally enhanced with respect to the noise. The variance
of the output filtered image is given by:
\begin{equation} \label{eq:varianza}
\sigma^2=\int{d{\bf q}\mathbf{\Psi}^T{\bf P}\mathbf{\Psi}}=\alpha.
\end{equation}
If a source has a signal-to-noise ratio $\geq 5$ after the filtering,
we say that we have a detection at that position.

Since we are dealing with point sources, the profiles $\tau_{\nu}$ are
directly given by the PSFs of the instrument. The cross-power spectrum
$\mathbf{P}$ is not known a priori but can be inferred from the data
under the assumption that the point sources are sparse. The unknown
frequency dependence, $\mathbf{f}=[f_{\nu}]$, of a given source can be
parametrised as:
\begin{equation} \label{eq:dep_frec}
\frac{S_{\nu}}{S_0} = \left( \frac{\nu}{\nu_0} \right)^{\gamma},
\end{equation}
where $S_{\nu}$ and $S_0$ are the flux densities at a frequency $\nu$
and at the frequency of reference $\nu_0$. The spectral index $\gamma$
is a free parameter. As pointed out by \cite{lanz10}, each image is
filtered several times for different MMFs. These MMFs are identical
except for the spectral index $\gamma$. The test values of the
spectral index used in this work are $-3.5 \le \gamma \le 3.5$, with a
step of 0.05. As it was shown in the cited work, the signal-to-noise
(SNR) ratio of the detected source is maximal for the correct choice
of this parameter, and by construction (we want an \emph{efficient}
estimator of the flux density of the source), the uncertainty assigned
to the source is the square root of the variance expressed in
eq.~(\ref{eq:varianza}) for the correct value of $\gamma$. While the
parametrisation of eq.~(\ref{eq:dep_frec}) is perfectly adequate in
our case, any other parametrisation, even a non-functional description
of the vector $\mathbf{f}$ by means of its components, could be used
in other cases, e.g. when more than two frequencies are considered
simultaneously.

\section{Application to WMAP data}\label{datos}

As mentioned above, the WMAP V and W bands are particularly
interesting because only a small fraction of WMAP sources were
detected with SNR $\ge 5\sigma$ in these bands. We have used an
adapted version of the code described in \cite{NEWPS07},
\cite{massardi09} and in \cite{powps}, modified in order to handle two
frequencies simultaneously and to accommodate our MMF filtering. The
code reads in an input parameter file containing the specific
characteristics of the maps to be analysed as well as the patch size,
the pixel size and overlap among the patches to effectively cover the
$100\%$ of the sky. Then, it reads in the two input maps in FITS
format and extracts the patches to be analysed using the tangential
plane approximation implemented in the CPAC library
\footnote{http://astro.ic.ac.uk/$\sim$mortlock/cpack/}.  Each pair of
V and W patches is analysed simultaneously using the MMF, and a first
set of detections is produced. Next, the code iteratively explores
different values of the spectral index for each source, allowing for
the appearance of new possible detections. At the end, the code
gathers all the candidates and generates a list of detections above a
given SNR, converting the positions of the detected objects in the
plane to the sphere. We have used flat patches of $14.6^{\circ} \times
14.6^{\circ}$, each containing $128 \times 128$ pixels. The pixel area
is $6.871^{\prime} \times 6.871^{\prime}$, corresponding to the
HEALPix resolution parameter $N_{side} = 512$ (see \cite{healpix} for
more details). In total we have 371 patches, with significant overlaps
among them in order to avoid as much as possible border effects. As
described in \cite{massardi09}, we perform a two-step process, first
doing a blind search across the sky and then refining the analysis
obtaining new patches centred at the positions of the sources
identified in the blind step. In this way we further reduce any
projection and border effect that we could have.

When dealing with the V and especially the W WMAP bands it is
necessary to carefully characterise the properties of the beams, both
to build the optimal filters (eq.~\ref{eq:mmf}) and to obtain a good
photometry of the sources. WMAP beams, particularly at 94 GHz, are
highly non-Gaussian and non circularly symmetric. We also need to take
into account that for the W band the pixel size ($6.871'$) is
dangerously close to the minimum sampling of the FWHM required to
avoid aliasing, and that real sources are not necessarily located at
the geometrical centre of the pixels. We have dealt with these
systematic effects in the following way.

We used the beam transfer functions (circularly symmetric) provided by
WMAP\footnote{
  lambda.gsfc.nasa.gov/product/map/dr4/beam\_xfer\_get.cfm} to create
high resolution templates of the beams (equivalent to the HEALPix
$N_{side}=4096$) projected into the tangent plane.  Please note that the maps that we have 
used to perform our analysis have a pixel size corresponding to $N_{side}=512$. In addition, 
we must consider that sources are not necessarily located at the
geometrical centre of these pixels. Therefore, taking the beam
transfer functions, we divided each $N_{side}=512$ pixel in $8\times8$
subpixels ($N_{side}=4096$) using the following formula:
\begin{equation} \label{eq:beam_Legendre}
b_l=2\pi\int{b^S(\theta)P_l(cos{\theta})d({cos{\theta}})/\Omega_B},
\end{equation}
where we obtained the value of the beam template in any angular
position ($\theta$) from the beam transfer function using the Legendre
polynomials. Then we averaged the high-resolution beams over all the
$8\times8$ possible displacements of the source centre inside a
$N_{side}=512$ pixel. In order to obtain a well-estimated average of
the beams, we repeated this process 5000 times. These averaged beam
templates were then degraded to the same resolution of our maps
($N_{side}=512$). In this way we take into account, on average, the
effects of pixelisation and the possible offsets of the sources with
respect to the geometrical centre of the pixel.

A careful correction for the non-circularity of the beams would
require a detailed knowledge of the combination of beam orientations
for all scans through every position in the sky. Since this
information is not available to us, we only applied to the flux
densities an average correction factor that takes into account the
effective beam areas for each channel. The correction factors were
obtained using the self-calibration method based on SExtractor
\citep{Sextractor} AUTO photometry, described in \cite{gnuevo08}. In
that work, the NEWPS 3yr sources detected above the $5\sigma$ level by
SExtractor and with flux densities $S\geq1$ Jy were used for the
self-calibration. Only 3 W-band sources satisfied these requirements
and therefore the correction factor for this band could not be
reliably estimated. To overcome this problem we have applied
SExtractor to the W map at the positions of 296 sources observed by
the Australia Telescope Compact Array (ATCA) and the Very Large Array
(VLA). Fifteen of them were detected at $\ge 3 \sigma$ by
SExtractor. Since all the ATCA/VLA sources have been detected at a
very high significance level, we were confident that there were no
spurious detections in this sample. These sources were used to
recalculate the effective beam area at 94 GHz. We obtained $(2.50 \pm
0.09) \times10^{-5}$ sr, to be compared with the nominal value of
$2.07 \times 10^{-5}$ sr for the symmetrised beam. Note that all the 
ingredients described in this section must be introduced in the code
before the filtering process.

\begin{figure*}
\begin{center}
\includegraphics[width=19.0cm]{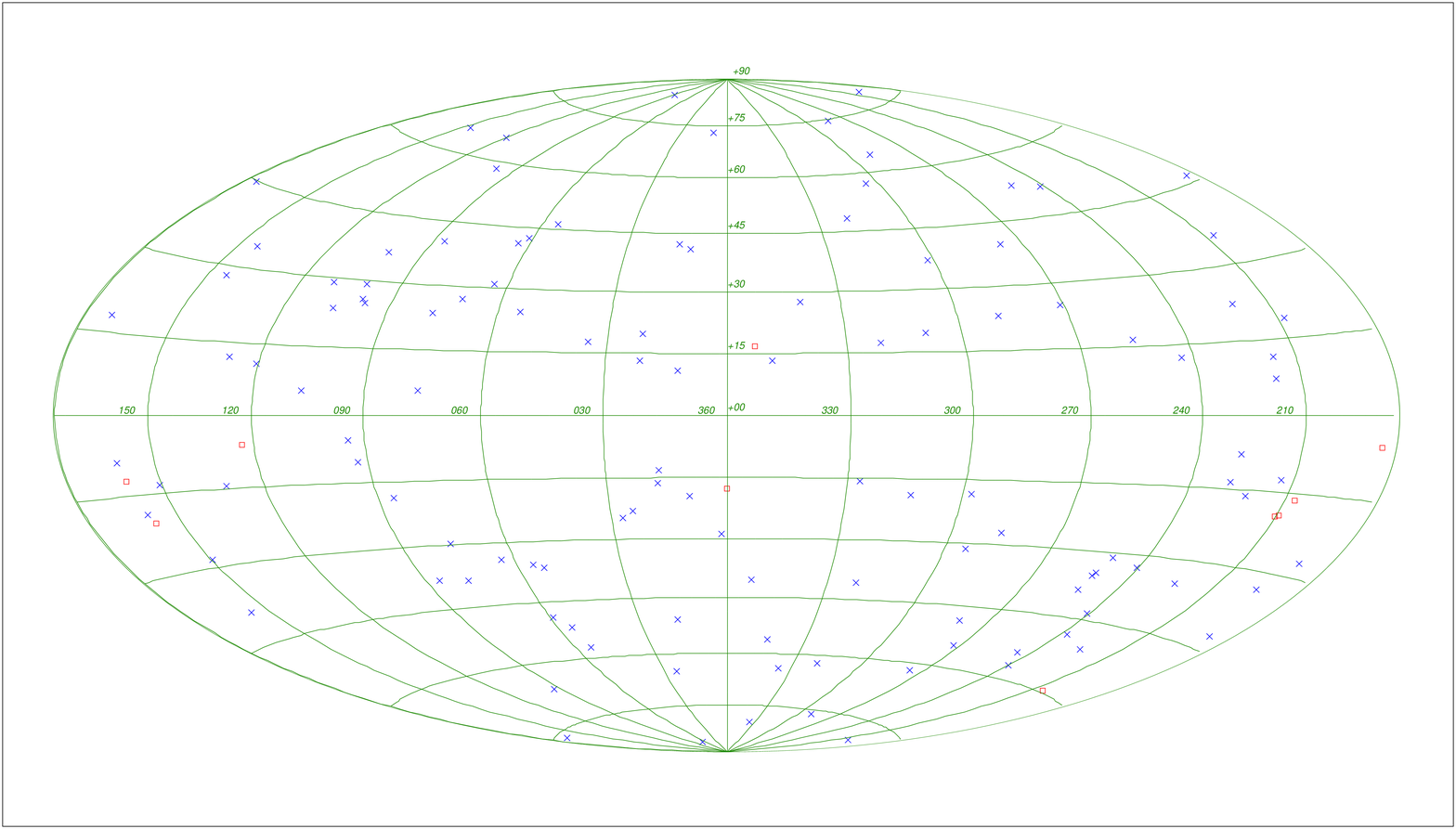}
\caption{Position on the sky, in Galactic coordinates, of the 129
  sources detected by the MMF on the at 61 and 94 GHz WMAP 7yr
  maps. Blue crosses represent the position of the 119 sources with
  confirmed counterparts in other catalogues (see
  Table~\ref{fuentes_extrag}); red squares represent the positions of
  the 10 Galactic or unidentified detections (see
  Table~\ref{fuentes_gal}). A coloured version of this figure is
  available with the online edition of the paper.}
\label{fuentes_cielo}
\end{center}
\end{figure*}
\twocolumn


\begin{table*}
\begin{center}
\begin{tabular}{rrrrrr}
\hline
RA ($^{\circ}$) & Declination ($^{\circ}$) & $S_{94\rm GHz}$ (Jy/beam) & rms (Jy) &  spectral index & Id in other catalogues \\
\hline
\hline

13.237 & 56.576 & 2.86 & 0.22 & -1.25 & NGC 281 \\
44.622 & -18.834 & 0.54 & 0.10 & -2.40 & $\cdots$ \\
52.264 & 31.338 & 4.81 & 0.57 & 4.50 & NGC 1333 \\
60.731 &  36.189 & 0.49 & 0.09 & -3.35 & NGC 1499 \\
83.654 & 22.045 & 155 & 2 & -1.15 & Crab Neb \\
83.807 & -5.419 & 216 & 3 & -0.35 & Orion region \\
84.018 & -6.338 & $\cdots$ & $\cdots$ & $\cdots$ & Orion region \\
85.396 & -1.917 & 30 & 2 & -0.2 & Orion region \\
246.756 & -24.633 & $\cdots$ & $\cdots$ & $\cdots$ & Ophiucus region \\
285.398 & -36.983 & 3.31 & 0.55 & 3.70 & NGC 6729 \\

\hline
\end{tabular}
\end{center}
\caption{Equatorial coordinates and identifications of the Galactic or
  unidentified detections. Sources without data are detections that, 
  due to the multifrequency method, do not have a clear estimation of 
  their flux density of the spectral index inside the studied range of 
  $\gamma$.}
\label{fuentes_gal}
\end{table*}

\onecolumn
\begin{center}
\begin{longtable}{rrrrrrr}
  \caption[Table]{Sources detected by the MMF on WMAP 7yr maps with
    counterparts in lower frequency catalogues. Presumably Galactic
    sources are listed in Table \ref{fuentes_gal}. The Equatorial
    coordinates (first two columns) are in degrees. The flags in the
    last column indicate if the source is listed in any of the the
    WMAP 7yr catalogues at any frequency (W) \citep{wmap7yr}, in
    NEWPS-5yr at 61 GHz (N) \citep{massardi09} or in the Planck
    catalogue at 70 or 100 GHz (P) \citep{ERCSC}. Sources without data
    are detections that, due to the multifrequency method, do not have
    a clear estimation of their flux density of the spectral index
    inside the studied range of $\gamma$. These sources with no flux
    density information are not taken into account in the subsequent
    analysis.}
\label{fuentes_extrag}\\
\hline \multicolumn{1}{r}{RA ($^{\circ}$)} &
\multicolumn{1}{r}{Declination ($^{\circ}$)} &
\multicolumn{1}{r}{$S_{94\rm GHz}$ (Jy/beam)} &
\multicolumn{1}{r}{rms (Jy)} &
\multicolumn{1}{r}{spectral index} &
\multicolumn{1}{r}{Id in other catalogues} &
\multicolumn{1}{r}{Flags} \\
\hline
\hline
\endfirsthead
\multicolumn{7}{r}%
{{\bfseries \tablename\ \thetable{} -- continued}} \\
\hline \multicolumn{1}{r}{RA ($^{\circ}$)} &
\multicolumn{1}{r}{Declination ($^{\circ}$)} &
\multicolumn{1}{r}{$S_{94\rm GHz}$ (Jy/beam)} &
\multicolumn{1}{r}{rms (Jy)} &
\multicolumn{1}{r}{spectral index} &
\multicolumn{1}{r}{Id in other catalogues} &
\multicolumn{1}{r}{Flags} \\
\hline
\hline
\endhead
\hline
\endfoot
\hline \hline
\endlastfoot
  1.483 &  -6.350 &   1.30 &   0.23 &  -0.80 &  PMN J0006-0623                  & WNP \\
  5.033 &  73.493 &   1.41 &   0.26 &  -0.45 &  GB6 J0019+7327                  & P   \\
  6.514 & -35.154 &   0.72 &   0.14 &  -1.65 &  PMN J0026-3512                  & WN  \\
  9.603 & -24.892 &   1.65 &   0.33 &   0.30 &  PMN J0038-2459                  & WNP \\
 16.715 & -40.600 &   1.74 &   0.21 &  -0.55 &  PMN J0106-4034                  & WNP \\
 23.160 & -16.928 &   0.91 &   0.17 &  -1.20 &  PKS 0130-17                     & WNP \\
 24.279 &  47.846 &   1.91 &   0.19 &  -1.20 &  GB6 J0136+4751                  & WNP \\
 24.383 & -24.518 &   1.56 &   0.30 &   0.20 &  PMN J0137-2430                  & WNP \\
 32.749 & -51.070 &   1.93 &   0.17 &  -1.00 &  PMN J0210-5101                  & WNP \\
 34.293 &  73.803 &   0.73 &   0.13 &  -1.70 &  GB6 J0217+7349                  & NP  \\
 39.510 &  28.836 &   2.20 &   0.33 &  -0.20 &  GB6 J0237+2848                  & WNP \\
 39.723 &  16.654 & $\cdots$ & $\cdots$ & $\cdots$ &  GB6 J0238+1637            & WP  \\
 43.374 & -54.666 &   1.56 &   0.20 &  -0.55 &  PMN J0253-5441                  & WNP \\
 46.088 & -62.233 &   1.31 &   0.25 &   0.05 &  PMN J0303-6211                  & WNP \\
 49.957 &  41.450 &   2.67 &   0.16 &  -1.90 &  GB6 J0319+4130                  & WNP \\
 50.377 & -37.092 &   0.81 &   0.14 &  -1.55 &  Fornax A                        & NP  \\
 53.549 & -40.203 &   1.47 &   0.25 &  -0.20 &  PMN J0334-4008                  & WNP \\
 54.252 &  32.327 &   2.57 &   0.42 &   0.70 &  GB6 J0336+3218                  & P   \\
 54.862 &  -1.734 &   2.33 &   0.33 &   0.05 &  PMN J0339-0146                  & WNP \\
 57.187 & -27.832 &   1.36 &   0.26 &   0.00 &  PMN J0348-2749                  & WNP \\
 61.062 & -36.120 &   2.14 &   0.21 &  -0.65 &  PMN J0403-3605                  & WNP \\
 64.548 &  38.025 &   1.96 &   0.16 &  -2.15 &  3C 111                          & NP  \\
 65.845 &  -1.374 &   3.39 &   0.25 &  -1.15 &  PMN J0423-0120                  & WNP \\
 67.166 & -37.869 &   1.66 &   0.29 &   0.05 &  PMN J0428-3756                  & WNP \\
 68.301 &   5.418 &   3.51 &   0.41 &   0.50 &  GB6 J0433+0521                  & WNP \\
 72.260 & -81.053 &   1.61 &   0.28 &   0.25 &  PMN J0450-8100                  & WNP \\
 73.914 & -46.248 &   3.05 &   0.26 &  -0.50 &  PMN J0455-4616                  & WNP \\
 74.270 & -23.427 &   1.66 &   0.20 &  -0.75 &  PMN J0457-2324                  & WNP \\
 78.849 & -45.994 &   1.02 &   0.19 &  -1.40 &  PMN J0515-4556                  & WP  \\
 79.975 & -45.795 &   2.30 &   0.21 &  -1.20 &  PMN J0519-4546                  & WNP \\
 80.849 & -36.456 &   2.57 &   0.19 &  -1.05 &  PMN J0522-3628                  & WNP \\
 84.678 & -44.028 &   3.15 &   0.21 &  -1.15 &  PMN J0538-4405                  & WNP \\
 91.968 &  -6.396 &   7.99 &   0.49 &   0.40 &  PKS 0605-06                     & NP  \\
 92.460 & -15.756 &   1.79 &   0.24 &  -0.50 &  PMN J0609-1542                  & WNP \\
 97.363 & -19.946 &   1.66 &   0.32 &   0.55 &  PMN J0629-1959                  & WNP \\
 98.985 & -75.242 &   2.03 &   0.19 &  -0.80 &  PMN J0635-7516                  & WNP \\
101.703 &  44.896 &   1.80 &   0.33 &   0.05 &  GB6 J0646+4451                  & WNP \\
102.564 & -16.699 &   1.75 &   0.26 &  -0.35 &  PKS 0648-16                     & NP  \\
110.556 &  71.313 &   2.15 &   0.26 &  -0.05 &  GB6 J0721+7120                  & WNP \\
111.493 &  -0.972 &   2.23 &   0.38 &   0.65 &  PMN J0725-0054                  & WP  \\
114.856 &   1.724 &   0.96 &   0.18 &  -1.35 &  GB6 J0739+0136                  & WNP \\
117.689 &  12.478 &   2.04 &   0.23 &  -0.90 &  GB6 J0750+1231                  & WNP \\
126.524 &   3.272 &   1.27 &   0.22 &  -0.90 &  GB6 J0825+0309                  & WNP \\
129.076 & -20.280 &   1.07 &   0.19 &  -1.10 &  PMN J0836-2017                  & WNP \\
130.356 &  70.885 &   0.91 &   0.15 &  -1.30 &  GB6 J0841+7053                  & WNP \\
133.730 &  20.128 &   2.48 &   0.29 &  -0.70 &  GB6 J0854+2006                  & WNP \\
140.233 &  44.738 &   0.89 &   0.14 &  -2.00 &  GB6 J0920+4441                  & WP  \\
140.441 & -26.324 &   1.07 &   0.21 &  -0.85 &  PMN J0921-2618                  & WP  \\
141.810 &  38.974 &   2.63 &   0.21 &  -1.15 &  GB6 J0927+3902                  & WNP \\
159.372 & -29.584 &   1.29 &   0.22 &  -0.75 &  PMN J1037-2934                  & WNP \\
159.659 &   5.136 &   0.80 &   0.16 &  -1.70 &  GB6 J1038+0512                  & W   \\
164.323 & -80.083 &   1.65 &   0.19 &  -0.85 &  PMN J1058-8003                  & WNP \\
164.615 &   1.562 &   2.86 &   0.22 &  -1.10 &  GB6 J1058+0133                  & WNP \\
171.870 & -18.836 &   0.97 &   0.19 &  -1.20 &  PMN J1127-1857                  & WP  \\
176.717 & -38.190 &   0.92 &   0.15 &  -1.75 &  PMN J1147-3812                  & WNP \\
178.285 &  49.483 &   1.47 &   0.19 &  -0.70 &  GB6 J1153+4931                  & WNP \\
179.831 &  29.192 &   1.21 &   0.22 &  -0.60 &  GB6 J1159+2914                  & WNP \\
187.270 &   2.104 &  10.71 &   0.33 &  -1.20 &  GB6 J1229+0202                  & WNP \\
187.699 &  12.338 &   5.99 &   0.26 &  -1.05 &  GB6 J1230+1223                  & WNP \\
191.732 & -25.848 & $\cdots$ & $\cdots$ & $\cdots$ &  PKS 1244-255              & WNP \\
193.992 &  -5.821 &   8.59 &   0.32 &  -1.15 &  PMN J1256-0547                  & WNP \\
194.823 &  51.645 &   1.37 &   0.25 &   0.05 &  GB6 J1259+5141                  & W   \\
197.630 &  32.472 &   0.82 &   0.13 &  -1.75 &  GB6 J1310+3220                  & WNP \\
201.390 & -42.960 &  16.92 &   0.46 &  -1.20 &  Centaurus A                     & NP  \\
204.426 & -13.021 &   4.96 &   0.29 &  -0.60 &  PMN J1337-1257                  & WNP \\
209.267 &  19.351 &   1.39 &   0.21 &  -0.75 &  GB6 J1357+1919                  & WNP \\
214.876 &  54.421 &   1.05 &   0.20 &  -0.50 &  GB6 J1419+5423                  & WNP \\
216.967 & -42.070 &   1.75 &   0.23 &  -0.75 &  PMN J1427-4206                  & WNP \\
229.382 & -24.370 &   2.31 &   0.28 &  -0.40 &  PMN J1517-2422                  & WNP \\
237.415 &   2.563 &   1.54 &   0.27 &  -0.55 &  GB6 J1549+0237                  & WNP \\
237.763 &   5.461 &   1.77 &   0.26 &  -0.50 &  GB6 J1550+0527                  & WNP \\
243.420 &  34.145 &   2.58 &   0.25 &  -0.30 &  GB6 J1613+3412                  & WNP \\
244.386 & -77.279 &   1.07 &   0.16 &  -1.20 &  PMN J1617-7717                  & WNP \\
246.561 & -29.832 &   1.91 &   0.30 &  -0.40 &  PKS 1622-29                     & NP  \\
248.816 &  38.155 &   3.25 &   0.25 &  -0.50 &  GB6 J1635+3808                  & WNP \\
249.747 &  57.392 &   1.98 &   0.29 &   0.55 &  GB6 J1638+5720                  & WNP \\
250.551 &  68.974 &   0.80 &   0.08 &  -2.45 &  GB6 J1642+6856                  & WNP \\
250.742 &  39.825 &   3.89 &   0.23 &  -0.75 &  GB6 J1642+3948                  & WNP \\
260.115 &  -1.025 &   2.18 &   0.27 &  -0.40 &  PKS 1717-00                     & NP  \\
260.873 & -65.025 &   1.32 &   0.25 &  -0.25 &  PMN J1723-6500                  & WP  \\
263.303 & -13.051 &   3.05 &   0.29 &  -0.30 &  PKS 1730-13                     & NP  \\
263.593 &  38.888 &   1.08 &   0.20 &  -0.70 &  GB6 J1734+3857                  & WNP \\
265.933 &  -3.813 &   2.55 &   0.22 &  -0.90 &  PKS 1741-03                     & NP  \\
267.966 &   9.655 &   4.05 &   0.26 &  -0.45 &  GB6 J1751+0938                  & WNP \\
268.340 &  28.738 &   1.65 &   0.21 &  -0.50 &  GB6 J1753+2847                  & WNP \\
270.301 &  78.459 &   1.05 &   0.17 &  -1.00 &  NVSS J180045+782804             & WP  \\
270.400 &  44.088 &   1.40 &   0.22 &  -0.45 &  GB6 J1801+4404                  & WNP \\
271.711 &  69.794 &   1.34 &   0.20 &  -0.25 &  GB6 J1806+6949                  & WNP \\
277.443 &  48.800 &   1.07 &   0.13 &  -1.60 &  GB6 J1829+4844                  & WNP \\
280.720 &  68.162 &   0.82 &   0.14 &  -0.90 &  GB6 J1842+6809                  & WP  \\
282.241 &  67.063 &   0.84 &   0.13 &  -1.15 &  GB6 J1849+6705                  & WNP \\
287.737 & -20.060 &   2.12 &   0.34 &   0.10 &  PMN J1911-2006                  & WNP \\
290.987 & -21.081 &   1.50 &   0.25 &  -0.65 &  PMN J1923-2104                  & WNP \\
291.214 & -29.282 &   7.01 &   0.28 &  -1.00 &  PMN J1924-2914                  & WNP \\
292.018 &  73.976 &   1.56 &   0.15 &  -1.20 &  GB6 J1927+7357                  & WNP \\
299.461 & -38.755 &   2.04 &   0.26 &  -0.55 &  PMN J1957-3845                  & WNP \\
299.942 &  40.752 &   6.02 &   0.36 &  -2.25 &  Cygnus A                        & NP \\
300.337 & -17.769 &   0.84 &   0.16 &  -1.60 &  PMN J2000-1748                  & WNP \\
302.849 & -15.774 &   1.45 &   0.29 &  -0.25 &  PMN J2011-1546                  & WNP  \\
314.018 & -47.226 &   1.87 &   0.25 &  -0.45 &  PMN J2056-4714                  & WNP \\
323.562 &  -1.791 &   1.25 &   0.22 &  -1.00 &  PMN J2134-0153                  & WNP \\
324.128 &   0.682 &   1.67 &   0.28 &  -0.30 &  GB6 J2136+0041                  & WNP \\
327.069 &   6.999 &   3.98 &   0.23 &  -1.15 &  GB6 J2148+0657                  & WNP \\
327.945 & -30.455 &   1.44 &   0.24 &  -0.55 &  PMN J2151-3028                  & WNP \\
328.366 &  47.275 &   1.93 &   0.26 &  -0.25 &  GB6 J2153+4716                  & NP  \\
329.184 & -69.703 &   1.81 &   0.22 &  -0.50 &  PMN J2157-6941                  & WNP \\
330.649 &  42.251 &   2.68 &   0.25 &  -0.30 &  GB6 J2202+4216                  & WNP \\
330.745 &  31.724 &   1.66 &   0.28 &  -0.05 &  GB6 J2203+3145                  & WNP \\
330.851 &  17.436 &   1.24 &   0.23 &  -0.70 &  GB6 J2203+1725                  & WNP \\
334.914 &  63.283 &   2.98 &   0.57 &   2.90 &  GB6 J2219+6317                  & NP  \\
336.405 &  -5.002 &   3.43 &   0.26 &  -0.75 &  PMN J2225-0457                  & WNP \\
337.427 &  -8.582 &   2.69 &   0.30 &  -0.35 &  PMN J2229-0832                  & WNP \\
338.171 &  11.674 &   4.19 &   0.34 &   0.20 &  GB6 J2232+1143                  & WNP \\
338.909 & -48.584 &   1.33 &   0.19 &  -0.85 &  PMN J2235-4835                  & WNP \\
341.600 & -12.087 &   1.24 &   0.21 &  -1.05 &  PMN J2246-1206                  & WP  \\
343.547 &  16.149 &   7.96 &   0.35 &  -0.35 &  GB6 J2253+1608                  & WNP \\
344.514 & -28.003 &   3.58 &   0.28 &  -0.35 &  PMN J2258-2758                  & WNP \\
352.347 & -47.438 &   1.18 &   0.22 &  -0.55 &  PMN J2329-4730                  & WP  \\
359.607 & -53.218 &   1.24 &   0.20 &  -0.60 &  PMN J2357-5311                  & WP  \\
\end{longtable}
\end{center}
\twocolumn

\section{Results and discussion} \label{resultados}

The source detection was performed applying the MMF filter to the WMAP
7yr V and W maps, at 61 and 94 GHz respectively. We have used only one
(V1) of the 2 differencing assemblies for the V band, because V1 and
V2 are very similar. As for the W band, we used 2 (W2 and W3) out of
the 4 differencing assemblies because their symmetrized beam profiles
are geometrically very similar (same as for W1 and W4) and this makes
the photometry easier. The W2 and W3 maps were combined pixel by
pixel, weighting with the inverse of the variances of the pixels. To
minimise the spurious detections due to the complex structure of the
Galactic emissions near the equatorial plane and of the Large
Magellanic Cloud (LMC) region, we have removed from the final
catalogue objects with $|b|\le 5^\circ$ and within a radius of
$5^\circ$ around the LMC ($(80.894^\circ,-69.756^\circ)$ in equatorial
coordinates). No sources are detected within the Small Magellanic
Cloud.

\subsection{Point source detection} \label{deteccion}

The two-step filtering approach described in section \ref{datos}
yields 129 $5\sigma$ detections in the 61 and 94 GHz WMAP 7yr maps,
outside the Galactic plane and LMC regions specified above. For each
of them, the MMF gives the flux density at the reference frequency (94
GHz, $\nu_0$ in eq. (~\ref{eq:dep_frec})) and the spectral index, the
only free parameter of the method. The uncertainty on the spectral
index can be estimated from intensive simulations made in
\cite{lanz10}. Table \ref{fuentes_gal} lists the nine sources that are
either well known Galactic sources (like the Crab Nebula, the HII
regions NGC~281 and NGC~1499 or the star-formation regions NGC~1333
and NGC~6729) or are located within Galactic molecular cloud
complexes, like the Orion or the Ophiucus regions, or within regions
of intense Galactic emissions, plus one unidentified. Table
\ref{fuentes_gal} gives the flux densities within the WMAP 94 GHz
beam. In most cases, these are lower limits to the total flux
densities because the sources are resolved. Given the different beam
sizes at the two frequencies, the spectral indices cannot be reliably
estimated in the case of resolved sources.

Excluding the ten sources listed in Table \ref{fuentes_gal}, the MMF
has detected 119 objects, listed in Table~\ref{fuentes_extrag}. All
these objects have a counterpart in lower frequency radio catalogues,
that cover the whole sky to much fainter flux density
limits\footnote{Except for a few of the sources in Table
  \ref{fuentes_gal}, none of the objects detected by the MMF has an
  IRAS counterpart.}, (see Figure~\ref{fuentes_cielo}). The apparently
ultra-steep ($\gamma_{61}^{94}= -2.25$) spectrum of Cygnus A is partly
due to resolution effects (the 94 GHz beam encompasses a lower
fraction of the total flux density than the 61 GHz beam).

In Tables~\ref{fuentes_gal} and \ref{fuentes_extrag}, sources without
an estimation of the flux density or the spectral index are objects
that are only `detected' at one frequency. Therefore, because the
method we use to filter the maps is multifrequency, the spectral
indices of these sources fall outside the range studied in this
work. These sources are not taken into account in the subsequent
analysis.

\begin{figure}
\begin{center}
\includegraphics[width=9.5cm] {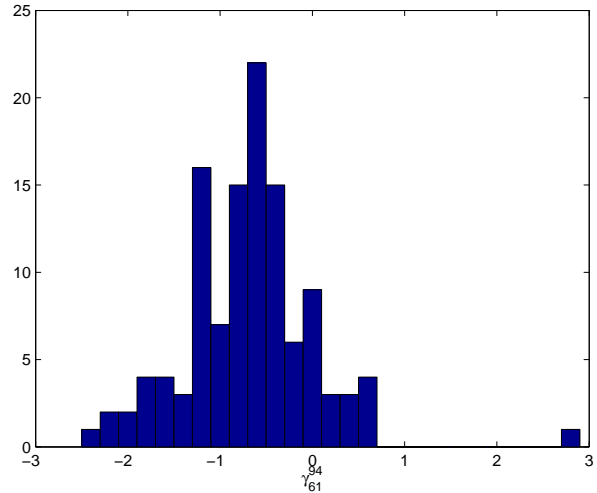}
\caption{Spectral index distribution of sources in
  Table~\protect\ref{fuentes_extrag}.
  \label{histograma}}
\end{center}
\end{figure}

The spectral index distribution of the sources in
Table~\ref{fuentes_extrag}, shown in Figure~\ref{histograma}, has a
median value of -0.65 with a dispersion of 0.71. For comparison,
\cite{wright09short}, mostly based on lower frequency WMAP data, found
a mean spectral index $\langle\gamma\rangle = -0.09$, with a
dispersion $\sigma=0.28$. Our results are therefore consistent with
the steepening of source spectra above $\sim 70\,$GHz found for
different data sets
\citep{sadler08,gnuevo08,marriage2011short,massardi2010,ERCSCprop}.


\begin{figure}
\begin{center}
\includegraphics[width=9.5cm] {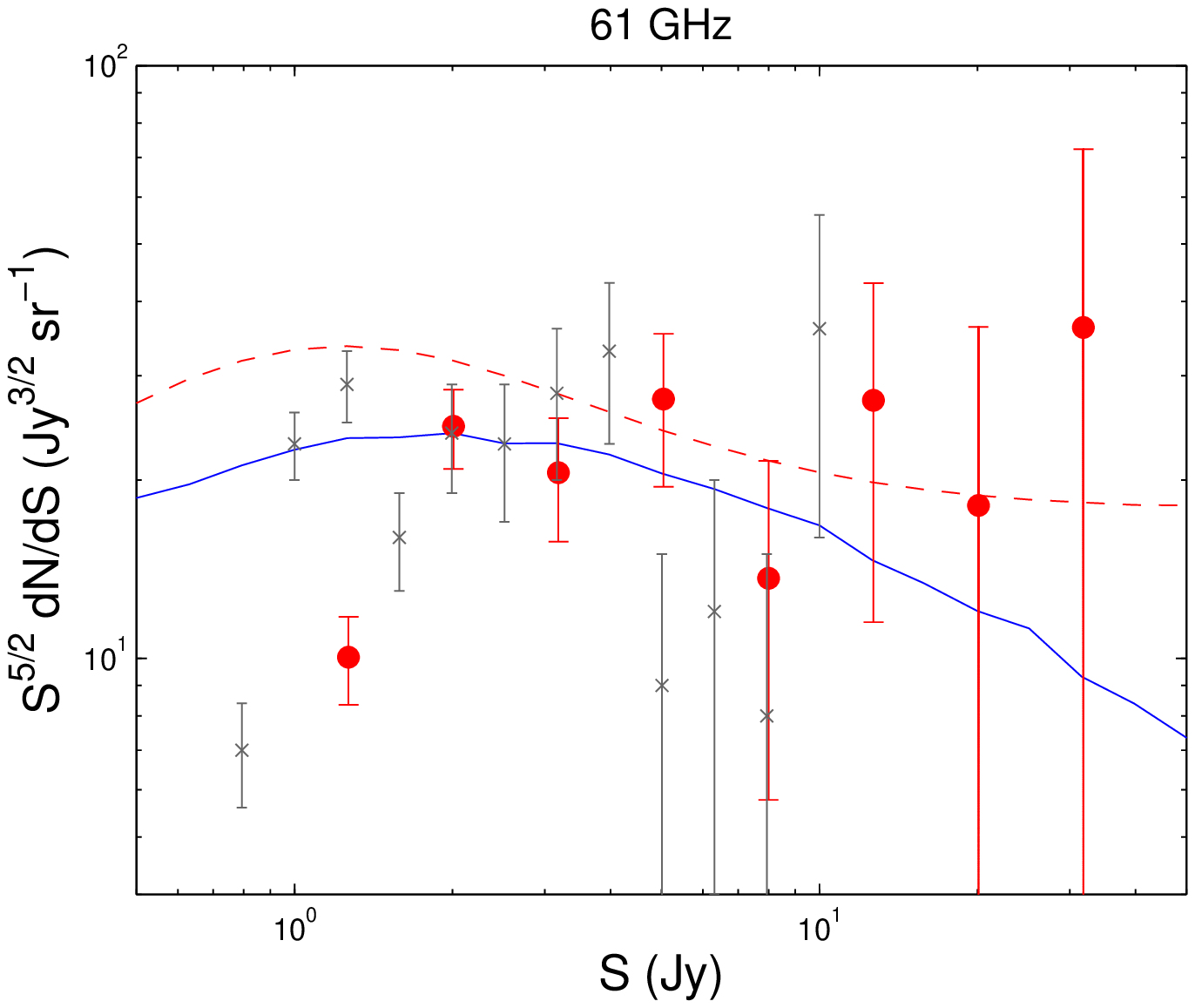}
\includegraphics[width=9.5cm] {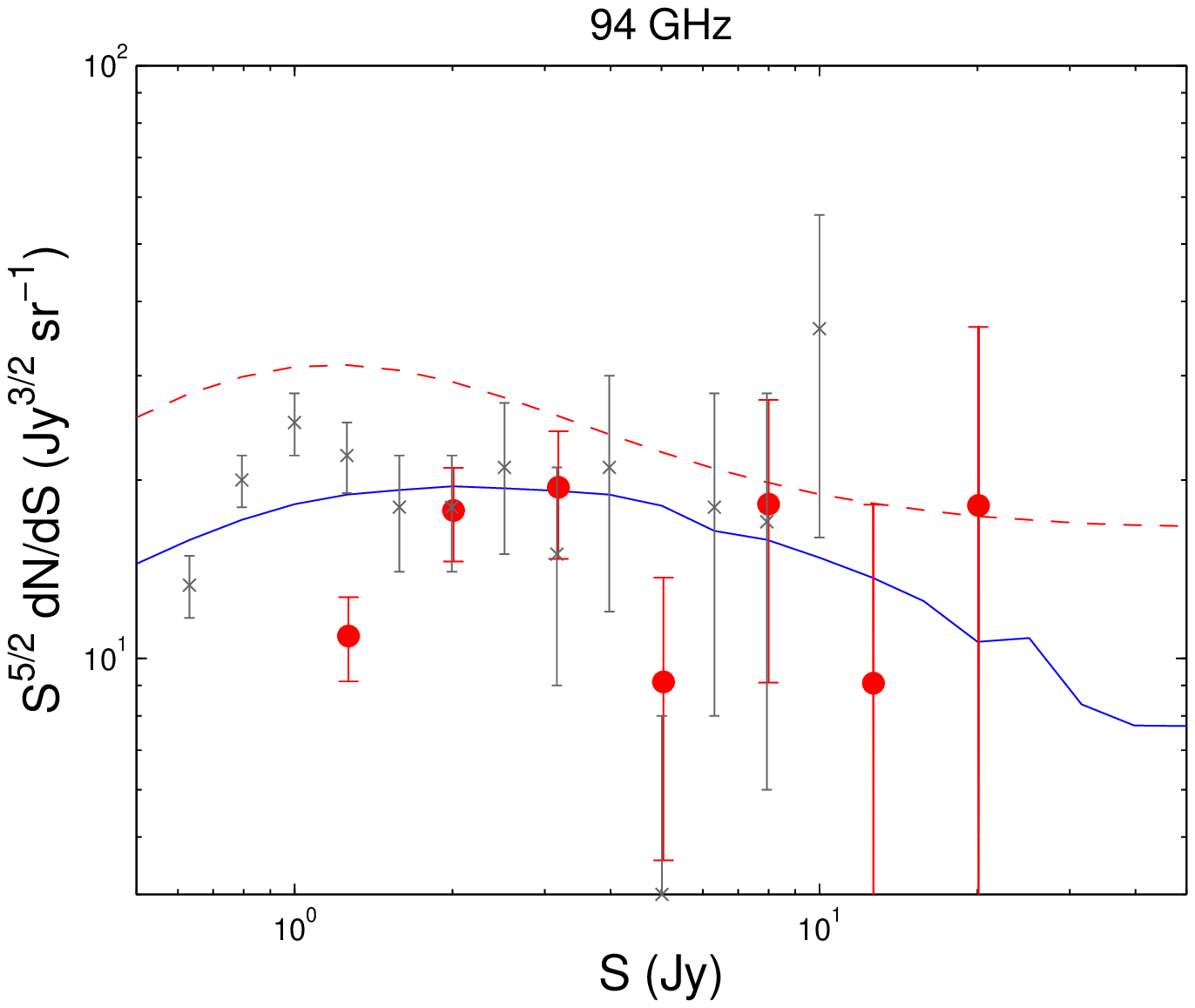}
\caption{WMAP 7yr Euclidean-normalised differential number counts at
  61 and 94 GHz based on the MMF sample (red points) with Poissonian
  error bars and the Bayesian correction to the flux densities. The
  grey crosses are the \emph{Planck} counts at 70 and 100 GHz
  \protect\citep{ERCSCprop}. The solid lines are the prediction by the
  \protect\cite{tucci11} model at 61 and 94 GHz and the dashed lines
  the prediction by the \protect\cite{zotti05} model at the same
  frequencies.}
\label{cuentas}
\end{center}
\end{figure}

\begin{table}
\begin{center}
\begin{tabular}{ccc}
\hline
\hline
S (Jy) & \multicolumn{2}{c}{$\hbox{S}^{5/2} \hbox{dN/dS}\,\hbox{(Jy}^{3/2}\,\hbox{sr}^{-1}\hbox{)}$} \\
\hline
       & 61\,GHz  & 94\,GHz \\
\hline
1.27 & 10.1$\pm$1.7 & 10.9$\pm$1.8 \\
2.01 & 25$\pm$4 & 18$\pm$3 \\
3.18 & 21$\pm$5 & 19$\pm$5 \\
5.05 & 27$\pm$8 & 9$\pm$5 \\
8.00 & 14$\pm$8 & 18$\pm$9 \\
12.68 & 27$\pm$16 & 9$\pm$9 \\
20.10 & 18$\pm$18 & 18$\pm$18 \\
31.85 & 36$\pm$36 & $\cdots$ \\
\hline
\end{tabular}
\end{center}
\caption{Euclidean-normalised differential number counts at 61 and 94
  GHz, based on our detections, in bins of $\Delta\log(S)=0.2$ and 
  taking into account the Bayesian correction to the flux densities.}
\label{tabla_cuentas}
\end{table}

The Euclidean normalised source counts at 61 and 94 GHz are given in
Table~\ref{tabla_cuentas} and shown in Figure~\ref{cuentas}, where
they are compared with the counts in the nearest \emph{Planck}
channels \citep{ERCSCprop} and with the predictions of the
\cite{zotti05} and \cite{tucci11} models.

In this figure, the flux densities obtained with the MMF were
corrected with a Bayesian approach \citep{BayesianCorrection} in order
to remove as much as possible the Eddington bias
\citep{eddingtonBias}. This approach takes into account the
distribution in flux density of the objects as a power law with
unknown slope, and an additive Gaussian noise. It is important to
point out that this correction is statistical, and therefore it has
been taken into account only in the estimation of the source counts.

The comparison in Figure~\ref{cuentas} shows that our completeness
limit is $\simeq 2\,$Jy. Above this limit, the agreement with the
\emph{Planck} counts and with the \cite{tucci11} model is generally
good. This confirms that the \cite{tucci11} model deals appropriately
with the high frequency behaviour of source spectra.

Although several data sets had suggested
\citep{sadler08,gnuevo08} and then detected \citep{ERCSCprop,planck_k}
a break in the bright extragalactic radio sources (at high-flux level)
at $\sim 70$ GHz, this break is only well explained with the current
data if we consider the \cite{tucci11} model.

\subsection{Comparison with the WMAP catalogues} \label{WMAP}

The WMAP Five-Band Point Source Catalog \citep{wmap7yr} lists all the
sources that where detected at $\ge 5\sigma$ in at least one frequency
channel, outside a mask excluding sources in the Galactic plane and
Magellanic Cloud regions. Flux densities in the other channels are
reported if the signal is detected at a $>2\sigma$ significance. In
this catalogue, 94 GHz flux densities are reported for 236
sources. However it is not clear how many of them are detected at a
$\ge 5\sigma$ level at 94 GHz. Since the WMAP and NEWPS catalogues
have comparable statistical completeness and reliability properties,
on the basis of the NEWPS3yr catalogue (see section \ref{sect:NEWPS})
we may roughly estimate that the $\ge 5\sigma$ detections are 22. We
expect a similar or smaller number of sources with the same
significance in the less complete WMAP catalogue at the same
frequency. By contrast, our method guarantees that all the 119
(presumably) extragalactic sources, 104 of which are outside the WMAP
Point Source Catalog mask, are detected at $\ge 5\sigma$ at 94 GHz.

\subsection{Comparison with NEWPS catalogues} \label{sect:NEWPS}

\citet{NEWPS07} looked for sources in the WMAP 3-yr maps at the
positions of known 5 GHz sources (non-blind search) and reported 22
detections at 94 GHz (listed in the NEWPS3yr
catalogue). \citet{massardi09} combined a non-blind (based on the ATCA
20 GHz survey catalogue) and a blind search on WMAP 5-yr maps,
excluding the W (94 GHz) channel (NEWPS5yr catalogue). At 61 GHz they
detected 169 sources, 61 of which are not present in our catalogue (if
we also take into account Table~\ref{fuentes_gal}); 45 of them however
are detected at $\ge 3\sigma$ with the MMF in the blind step of our
detection process. Seven of the other 18 sources are at $|b| <
15^{\circ}$ and may be contaminated by Galactic emission, and three
lie in the LMC region, excluded from our search. The remaining 8
sources have not been detected by the MMF. On the other hand, our
catalogue of extragalactic objects contains 16 sources not present in
the NEWPS5yr catalogue. We must remark that our catalogue is
completely based on a blind search, whereas the NEWPS catalogue was
guided by a previous selection of sourecs detected at low frequency.

\subsection{Comparison with ATCA and NRAO flux densities} \label{sect:ATCA}

\begin{figure}
\begin{center}
\includegraphics[width=9.5cm] {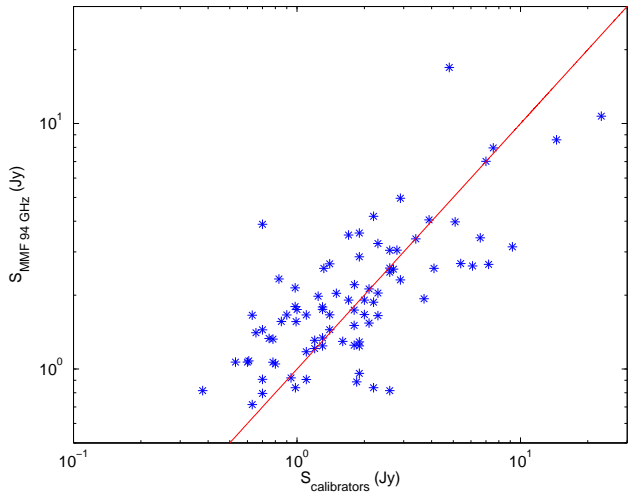}
\caption{Comparison between 94 GHz flux densities recovered by the MMF
  and the ATCA (3 mm) and NRAO (90 GHz) flux density measurements, for
  the 85 sources in common. No correction for the slightly different
  frequencies of the sets of measurements was made. The solid line is
  x=y.
  \label{ATCA}}
\end{center}
\end{figure}

Our sample of 94 GHz detections includes 85 sources with ground based
$\simeq 3$ mm observations either with
ATCA\footnote{http://www.narrabri.atnf.csiro.au/calibrators/index.html}
or with the NRAO 12m telescope
\citep{Holdaway1994}\footnote{http://www.alma.nrao.edu/memos/html-memos/alma123/memo123.pdf}. The
ATCA flux densities for these sources have been collected in the
frequency range between 85 and 105 GHz with the ``old'' ATCA digital
correlator with up to $2\times256$ MHz
bandwidth. 
The MMF flux density estimates are compared with ATCA and NRAO
measurements in Figure~\ref{ATCA}. There is evidence that the MMF
somewhat overestimates the flux densities below $\simeq 2\,$Jy, most
likely due to the Eddington bias. Above 2 Jy the mean absolute
fractional error $\langle|S_{\rm MMF}-S_{\rm ground}|/S_{\rm
  ground}\rangle\simeq 38\%$ somewhat higher than expected from the
combination of nominal measurement errors and variability, suggesting
that the true errors associated to the MMF flux density estimates are
somewhat larger than the nominal values. In this respect, it must be
noted that the WMAP 7-yr maps are averages over seven years of
observations while the ATCA and NRAO measurements refer to a single
epoch.

\begin{figure}
\begin{center}
\includegraphics[width=9.5cm] {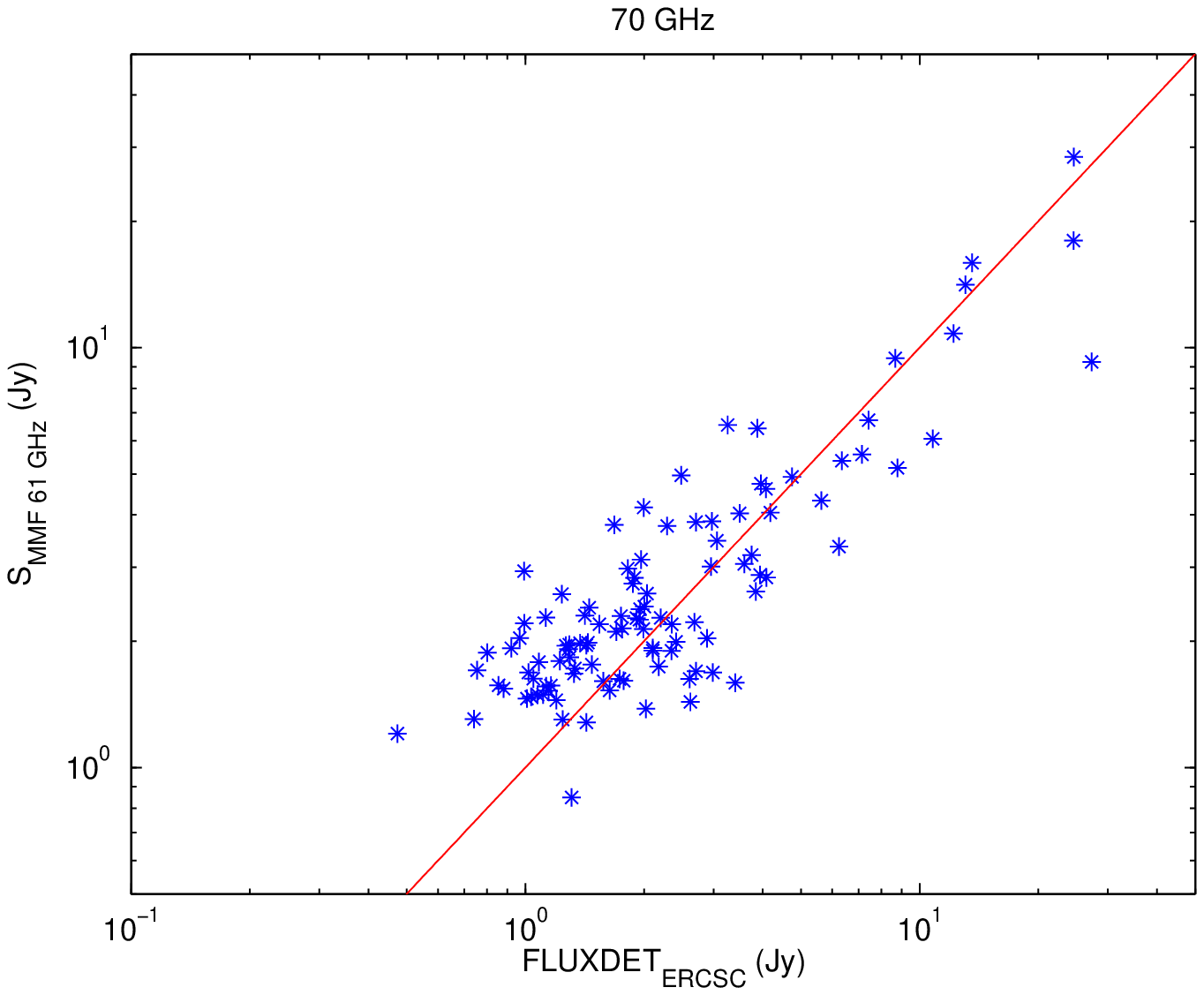}
\includegraphics[width=9.5cm] {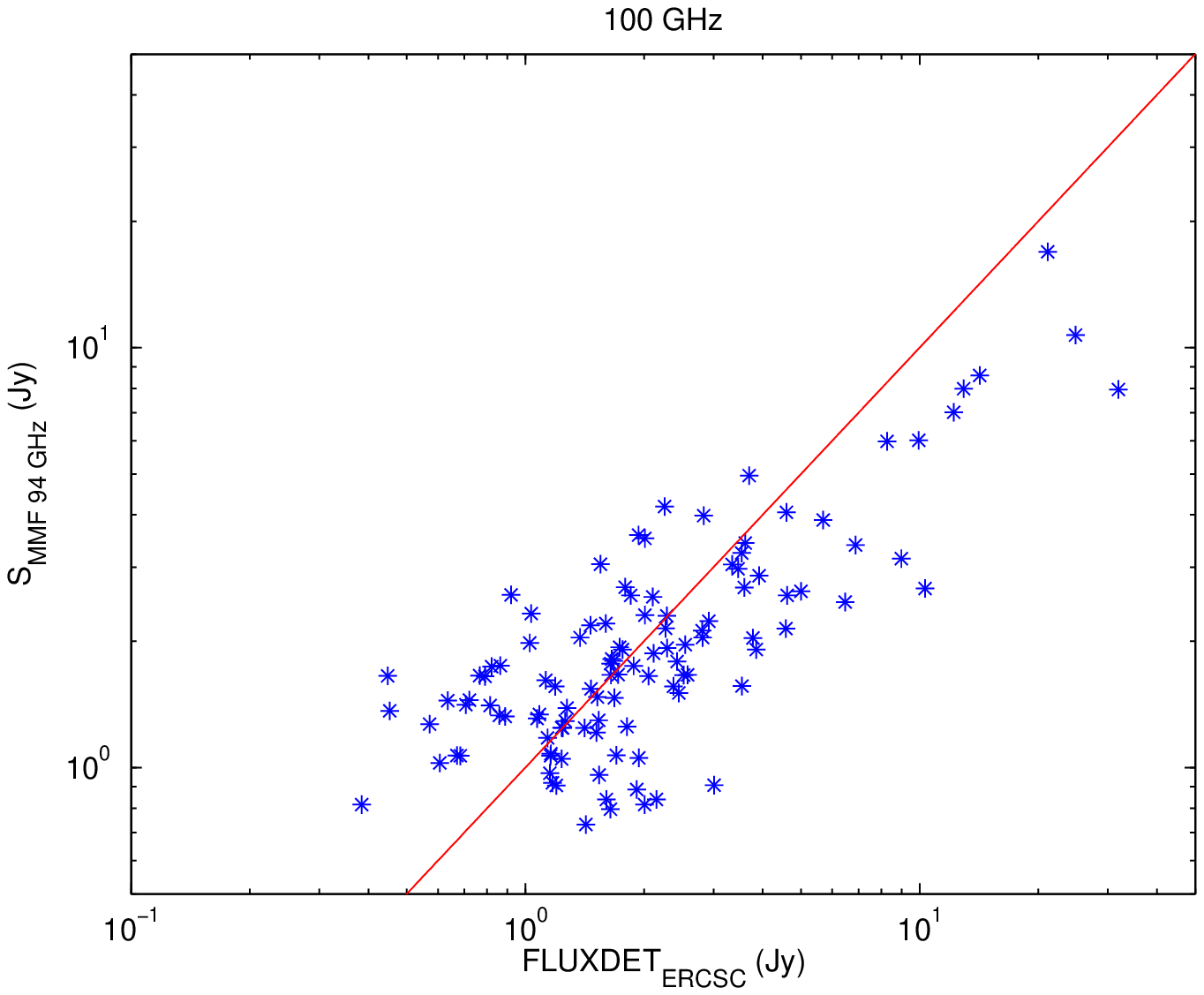}
\caption{Comparison of the flux densities recovered by the MMF on WMAP
  maps at 61 and 94 GHz with the FLUXDET values reported in the
  \emph{Planck} ERCSC at 70 GHz (upper panel) and 100 GHz (lower
  panel), respectively. The solid line is x=y.
  \label{planck}}
\end{center}
\end{figure}

\subsection{Comparison with the \emph{Planck} Early Release Compact Source Catalogue 
(ERCSC)} \label{ERCSC}

Thanks to its higher sensitivity, \emph{Planck} has detected far more
sources than WMAP. The Early Release Compact Source Catalogue (ERCSC)
lists 1381 sources detected at 100 GHz and 599 detected at 70 GHz; 308
and 788 of the ERCSC 70 GHz and 100 GHz sources, respectively, are
outside of the WMAP Point Source Catalog mask. Three of our sources in
Table~\ref{fuentes_extrag} are not present in the ERCSC, but are
present in lower frequency catalogues with flux densities consistent
with those inferred from WMAP data.

The ERCSC gives four different measures of flux density for each
source. For the comparison with our results, we have chosen the
estimate called FLUXDET, which is calculated using an approach similar
to ours and appears to have higher reliability for low signal-to-noise
ratios.
Figure~\ref{planck} compares the FLUXDET flux densities at 70 and 100
GHz with the MMF ones at the nearest WMAP frequencies (61 and 94 GHz,
respectively). Again there is evidence that the MMF flux densities are
affected by the Eddington bias below $\simeq 2\,$Jy. The MMF flux
densities of the 5 sources with $S_{\rm MMF, 94GHz}\gtrsim 8\,$Jy are
all lower than the the ERCSC flux densities at 100 GHz. Two of these
sources (the one at $\hbox{RA}= 91.968^\circ$,
$\hbox{dec}=-6.396^\circ$ and Cen A) has a strong dust emission, seen
in the IRAS survey, that enhances the 100 GHz flux density. The other
3 are well known highly variable blazars (3C273, 3C279, 3C454.3),
caught by Planck in a bright
phase. 

The spectral index distributions of our sources are pretty close to
that of the ERCSC sources: the median spectral index of our sources is
-0.65, with a standard deviation of 0.71; for ERCSC sources with
$\gamma \le 2$ (to avoid strong contamination by dust emission) and
excluding the Galactic plane and the LMC region as defined previously
in order to have a sample comparable with ours, we find a median value
of $-0.39$ with a standard deviation of 0.52.

\section{Conclusions} \label{conclusiones}

The detection of extragalactic point sources is a crucial task in the
analysis of CMB maps because point sources are the main contaminant of
the CMB power spectrum on small angular scales. From the same maps we
can extract interesting astrophysical information about the point
sources themselves. The development of efficient detection tools is
therefore very important. The MMF holds the promise of achieving
detections down to fainter flux densities than achievable by standard
methods.

Applying this tool to WMAP 7yr maps at 61 and 94 GHz simultaneously we
have obtained 129 $5\sigma$ detections at both frequencies, outside
the Galactic plane (i.e. at $|b|>5^\circ$) and the LMC regions. Nine
of these sources, listed in Table~\ref{fuentes_gal}, are either known
Galactic sources or reside in regions of high Galactic emission. One
additional source, also listed in Table~\ref{fuentes_gal}, does not
have a counterpart in low frequency radio catalogues. All the other
119 sources, listed in Table~\ref{fuentes_extrag}, do have a low
frequency counterpart; 104 of them reside outside the WMAP Point
Source Catalog mask. For comparison, the NEWPS-3year catalogue
contains 22 $5\sigma$ detections at 94 GHz. Although the \emph{Planck}
ERCSC contains far more sources at the frequencies (70 and 100 GHz)
nearest to those used in the present analysis, we have detected three
sources not present in the ERCSC, yet with flux densities consistent
with lower frequency measurements.

A comparison of the flux densities yielded by the MMF with those
obtained at $\simeq 90\,$GHz with the ATCA or the NRAO 12m telescope,
and with the \emph{Planck} ERCSC data at 70 and 100 GHz, shows a
generally good agreement above $\simeq 2\,$Jy, although the rms
differences between MMF and ground based or ERCSC values are larger
than expected from variability and nominal measurement errors. This
suggests that the errors associated to MMF flux density estimates are
somewhat larger than the nominal values listed in
Table~\ref{fuentes_extrag}. Below 2 Jy, the MMF flux densities tend to
be somewhat overestimated, as the effect of the Eddington bias.

The distribution of 61--94 GHz spectral indices of sources in
Table~\ref{fuentes_extrag} has a median value equal to
$\gamma_{61}^{94} = -0.65$ (with a dispersion of 0.71), while
\cite{wright09short}, mostly based on lower frequency WMAP data, found
a mean spectral index $\langle\gamma\rangle = -0.09$, with a
dispersion $\sigma=0.28$. Our results therefore confirm the steepening
of extragalactic radio source spectra above $\sim 70$ GHz, suggested
by various data sets
\citep{sadler08,gnuevo08,marriage2011short,massardi2010} and then
confirmed by analyses of different samples
\citep{ERCSCprop,planck_k,giommi11} of bright extragalactic sources
extracted by the Planck Early Release Compact Source Catalogue
\citep{ERCSC}.

This average steepening observed in the spectra of bright
extragalactic sources, which are mainly blazars in this frequency
range \citep{Angel-Stockman}, has been recently interpreted by
\cite{tucci11} in terms of the "break" frequency, foreseen in the
emission spectra of blazars by classic physical models for the
synchrotron emission in the inner part of inhomogenous conical
jets. Remarkably, this new model by \cite{tucci11} is able to give a
good a fit not only to the source number counts here presented (see
Figure~\ref{cuentas}; \S\,\ref{deteccion}) but also to almost all the
published data on number counts and spectral index distributions of
bright extragalctic radio sources up to, al least, 200-300 GHz.

\section*{Acknowledgments}

The authors acknowledge partial financial support from the Spanish
Ministerio de Econom\'ia y Competitividad projects
AYA2010-21766-C03-01 and CSD2010-00064. LFL acknowledges the Spanish
CSIC for a JAE-Predoc fellowship and the hospitality of the Scuola
Internazionale Superiore di Studi Avanzati (SISSA) of Trieste during
two research stays in 2010 and 2011. ML-C thanks the Spanish
Ministerio de Econom\'ia y Competitividad for a Juan de la Cierva
fellowship. JG-N, GdZ, and MM acknowledge financial support from
ASI/INAF Agreement I/072/09/0 for the Planck LFI activity of Phase
E2. DH wishes to acknowledge the Spanish Ministerio de Educaci\'on for
a Jos\'e Castillejo Fellowship (JC2010-0096) and the Astronomy Group
at the Cavendish Laboratory for their hospitality during a research
stay in 2011. The authors are grateful to Luigi Toffolatti and Marco
Tucci for their useful comments and explanations, in particular on the
extragalactic source counts.

\bibliographystyle{mn2e}
\bibliography{draft_bib}

\label{lastpage}

\end{document}